\documentclass[a4paper]{spie}
\usepackage[]{graphicx}
\usepackage{amsmath,amssymb}
\usepackage{bm}

\def\Ba133{${}^{133}$Ba}
\def\Co57{${}^{57}$Co}
\def\Cs137{${}^{137}$Cs}
\def\Am241{${}^{241}$Am}
\def\Na22{${}^{22}$Na}

\def\micron{$\mu$m}					
\def\degree{\kern-.2em\r{}\kern-.3em} 
\def\degC{\kern-.2em\r{}\kern-.3em C} 
\def\degreeC{\kern-.2em\r{}\kern-.3em C} 
\def\dC{\kern-.2em\r{}\kern-.3em C} 

\title{Results of a Si/CdTe Compton Telescope}

\author{Kousuke Oonuki\supit{a,b}, Takaaki Tanaka\supit{a,b}, Shin Watanabe\supit{a}, Shin'ichiro Takeda\supit{a,b}, 
\\
Kazuhiro Nakazawa\supit{a}, Takefumi Mitani\supit{a,b}, Tadayuki Takahashi\supit{a,b}, Hiroyasu Tajima\supit{c}, 
\\
Yasushi Fukazawa\supit{d}, Masaharu Nomachi\supit{e}
\skiplinehalf
\supit{a}Institute of Space and Astronautical Science (ISAS/JAXA), Sagamihara, Kanagawa 229-8510, Japan\\
\supit{b}Department of Physics, University of Tokyo, Bunkyo, Tokyo 113-0033, Japan
\\
\supit{c}Stanford Linear Accelerator Center, Stanford, CA 94309-4349
\\
\supit{d}Department of Physics, Hiroshima University, Higashi-Hiroshima, 739-8526, Japan
\\
\supit{e}Department of Physics, Osaka University, Toyonaka, Osaka, 560-0043, Japan
}

\authorinfo{Further author information: (Send correspondence to K.O.)\\K.O.: E-mail: oonuki@astro.isas.jaxa.jp, Telephone: 81 42 759 8510}

\begin{document} 
 \maketitle 

\begin{abstract}
We have been developing a semiconductor Compton telescope to explore the universe in the energy band from several tens of keV to a few MeV. We use a Si strip and CdTe pixel detector for the Compton telescope to cover an energy range from 60 keV.
For energies above several hundred keV, the higher efficiency of CdTe semiconductor in comparison with Si is expected to play an important role as an absorber and a scatterer. In order to demonstrate the spectral and imaging capability of a CdTe-based Compton Telescope, we have developed a Compton telescope consisting of a stack of CdTe pixel detectors as a small scale prototype. With this prototype, we succeeded in reconstructing images and spectra by solving the Compton equation from 122 keV to 662 keV.  The energy resolution (FWHM) of reconstructed spectra is 7.3 keV at 511 keV and 3.1 keV at 122 keV, respectively.  The angular resolution obtained at 511 keV is measured to be 12.2\degree{}(FWHM).

\end{abstract}


\section{INTRODUCTION}
The hard X-ray and gamma-ray bands ranging from several tens of keV to a few MeV are important windows for exploring the energetic universe.  It is in these regions that high energy phenomena such as nucleosynthesis and particle acceleration become dominant. 
However, observation sensitivity in this energy band has been still limited, due to high background, low detection efficiency, and difficulty of imaging by means of focusing technology. A semiconductor Compton telescope utilizing the good energy and position resolution of semiconductors is one of the most promising ways to realize a breakthrough in this energy band. 

Based on our recent achievements for high resolution  CdTe and Si imaging detectors\cite{Ref:Takahashi_IEEE2002,Ref:Tanaka2,Ref:Nakazawa,Ref:Fukazawa,Ref:Tajima2}, 
we have proposed a Si/CdTe semiconductor Compton telescope as a next generation Compton telescope\cite{Ref:Takahashi_NeXT,Ref:Takahashi_SPIE}.  The basic concept of the Si/CdTe Compton telescope is to utilize Si as a scatterer and CdTe as an absorber.  
Si works as a good scatterer with energies higher than 60 keV due to large Compton scattering efficiency, and CdTe has high photo absorption efficiency as an absorber due to the large atomic numbers of Cd (Z = 48) and Te (Z = 52). The Si/CdTe Compton telescope surpasses scintillator-based Compton telescopes by its high energy and angular resolution. 
According to the measurements with our prototype which is composed of six layers of Double-sided Si strip detector (DSSD) with a thickness of 300 $\mu$m and CdTe pixel detectors with a thickness of 500 \micron, we have successfully achieved Compton reconstructed images of gamma-rays from 81~keV to 662~keV, and an angular resolution of 3.9 \degree{} for 511~keV gamma-rays\cite{Ref:Mitani,Ref:Tanaka,Ref:Watanabe_COMP}.

In order to achieve much higher efficiency for gamma-rays above a few hundred keV, we need a CdTe detector as thick as several mm. One idea to achieve such a thick device is a stacked CdTe detector \cite{Ref:Takahashi_NIM2000,Ref:Watanabe_IEEE2002,Ref:Watanabe_IEEE2003}.
At energies above 300 keV, the stacked CdTe detector by itself works as a Compton telescope.  To understand the behavior of the stacked detector, we developed a prototype, which is composed of three layers of CdTe pixel detectors along with one at their side. In this paper, the results obtained with the prototype are described.
The comparison with the results obtained from a Compton telescope consists of DSSD and CdTe pixel detectors are also carried out.

\section{A Si/CdTe Compton Telescope}

\begin{figure}[t]
\begin{center}
		\includegraphics[width = 7 cm, keepaspectratio, clip]{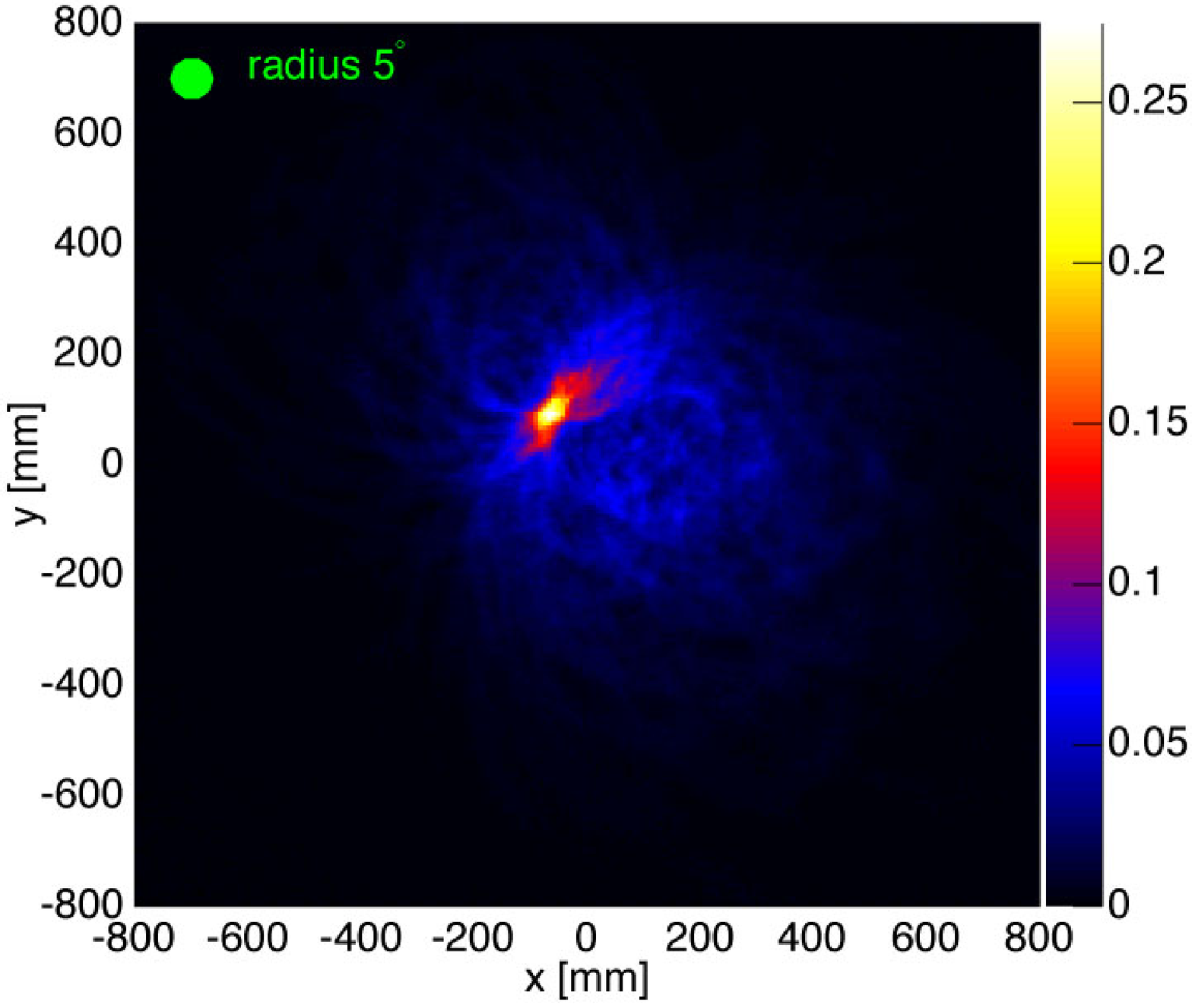}
\hspace{1cm}
		\includegraphics[width = 6cm, keepaspectratio, clip]{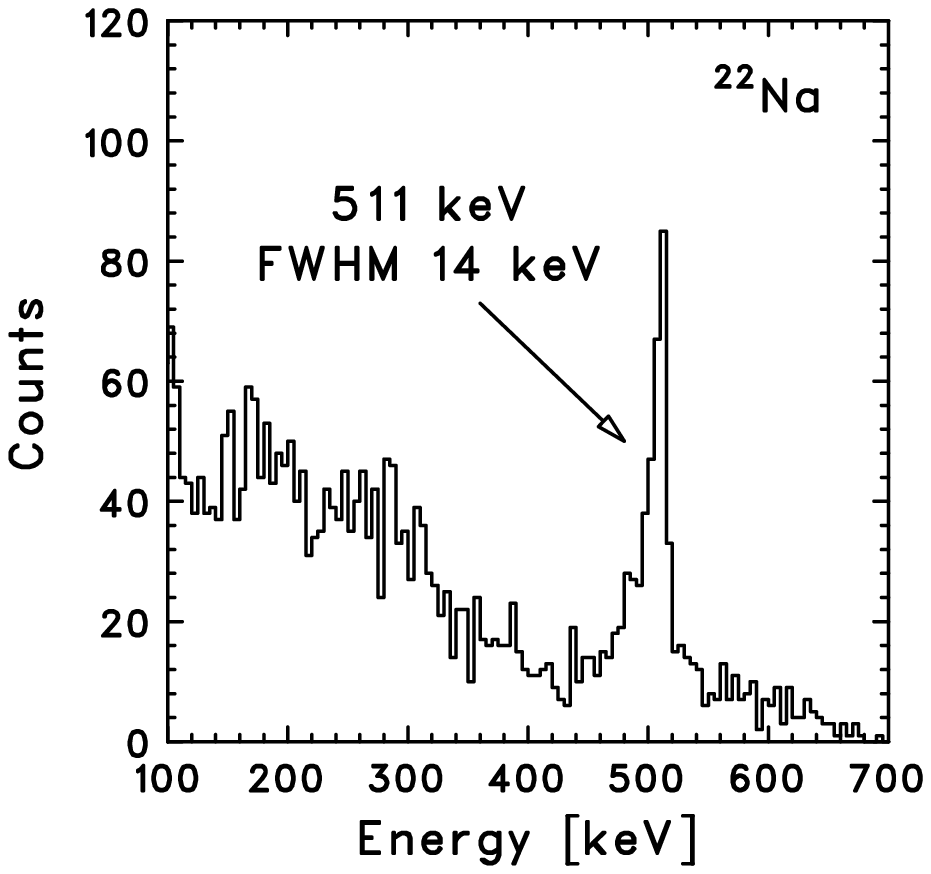}

		\caption {Compton reconstructed image (left) and spectrum (right) of the 511 keV gamma-ray obtained with a prototype Si/CdTe Compton telescope. The angular resolution is 3.9\degree{}.  The energy resolution is 14 keV (FWHM) at 511 keV.}
		\label{fig: Si/CdTe Image}
		\label{fig: Si/CdTe Spectrum}
\end{center}
\end{figure}

As described in our previous publications \cite{Ref:Mitani,Ref:Tanaka,Ref:Watanabe_COMP}, we have constructed a prototype Si/CdTe Compton telescope with different configurations. The high energy resolution of a newly developed DSSD has proven to be essential to obtain not only a wide energy coverage starting from 60 keV, but also a high angular resolution.  With the prototype Si/CdTe Compton telescope, we obtained Compton reconstructed images of the gamma-rays from 81 keV to 662 keV. The 511 keV image is shown in the Fig. \ref{fig: Si/CdTe Image}. The angular resolution is measured to be 3.9 \degree{} for 511 keV gamma-rays.  Fig. \ref{fig: Si/CdTe Spectrum} shows the reconstructed spectrum for 511 keV gamma-rays from a $^{22}$Na gamma-ray source. The energy resolution of 14 keV (FWHM) demonstrates the high potential of the Si/CdTe Compton telescope for future missions.

Since the thickness of the CdTe pixel detector adopted as an absorber in the prototype is only 0.5 mm, the total efficiency for gamma-rays above several hundred keV is very limited.  A possible improvement to this is to thicken the CdTe detector. A monolithic and thick detector is not, however, applicable if we use the current CdTe or CdZnTe semiconductor, due to incomplete charge collection.
Therefore, we are working on the concept of a stacked CdTe detector with a capability of imaging.


\section{A Stacked thin CdTe pixel detector for Compton Telescope}

For gamma-rays above a hundred keV, adequate detection efficiency can be obtained by CdTe or CdZnTe with a thickness of more than 5 mm.
It is, however, difficult to increase the thickness of the device, keeping good energy resolution throughout the energy range from several ten keV to a few hundred keV.
This is because the effect of incomplete charge collection becomes more severe due to small mobility and short lifetime of holes of CdTe or CdZnTe. The inactive region in the detector volume degrades the energy resolution through a low energy tail in the spectral response \cite{Ref:Takahashi_IEEE2001,Ref:Takahashi_IEEE2001_2}.

In order to achive fully active CdTe or CdZnTe detectors, the concept of using thin device is of particular importance, since we can apply sufficient bias voltage to collect all charge produced in the detector. As calculated in our paper \cite{Ref:Watanabe_IEEE2002}, a 0.5 mm thick CdTe detector becomes fully active with a bias voltage of 800 V, and even holes generated near the anode face can be completely collected.  By using a detector with a thickness of 0.5 mm, an applied bias voltage of 1400 V, and a temperature of $-$20 \degC{}, we have achieved an energy resolution (FWHM) of 830 eV at 59.5 keV, and 2.1 keV at 662 keV, which is close to the Fano limit of CdTe \cite{Ref:Takahashi_IEEE2001_2}.

\begin{figure}[t]
\begin{center}

	\begin{minipage}[t]{7cm}
	\begin{center}
		  \includegraphics[width = 6 cm, keepaspectratio, clip]{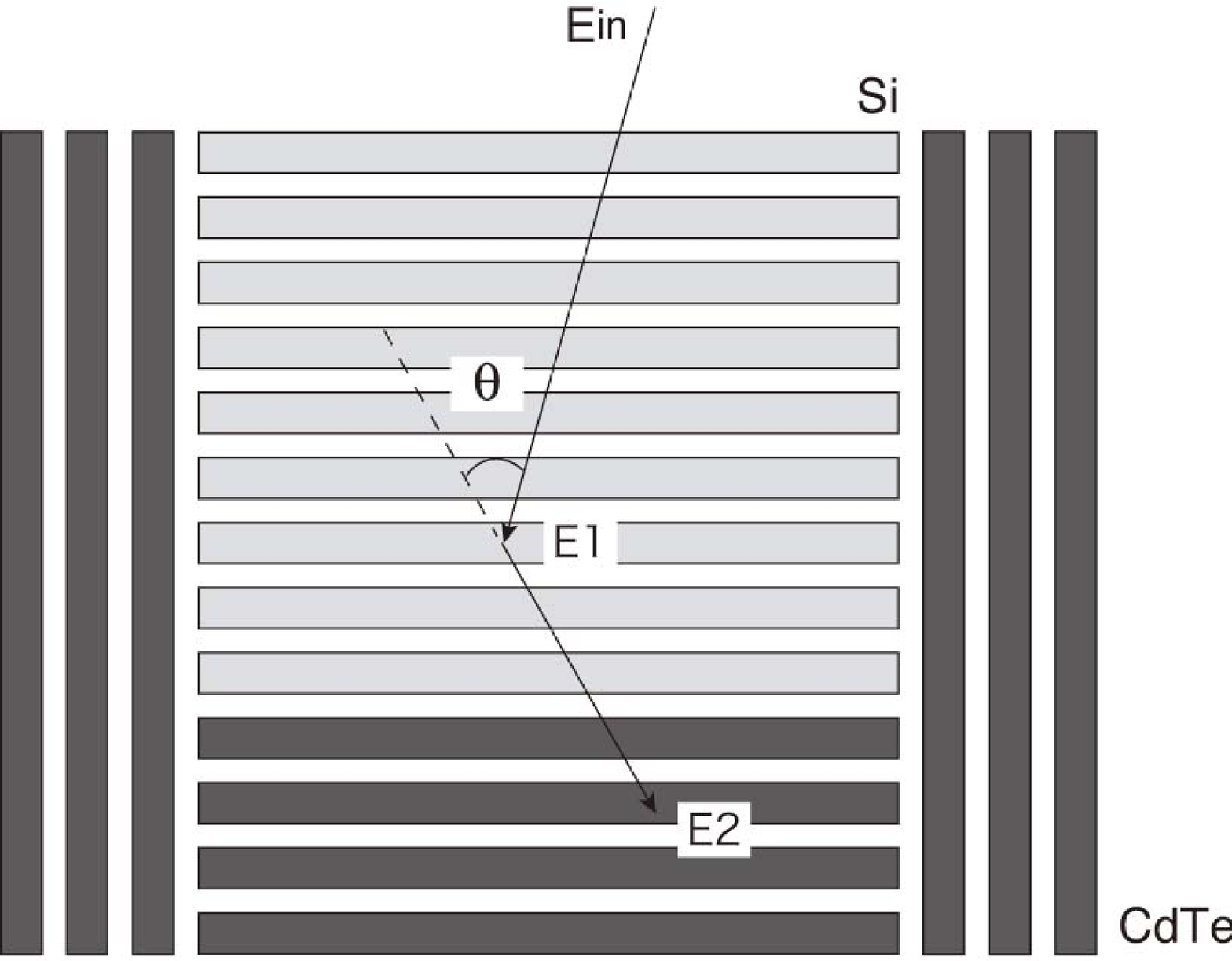}
	\end{center}
		\caption {Conceptual drawing of our Si/CdTe semiconductor Compton telescope. CdTe pixel detectors are stacked into 20 layers to achieve larger effective area.}
		\label{fig: Si/CdTe concept}
	\end{minipage}
\hspace{0.5cm}
	\begin{minipage}[t]{9cm}
	\begin{center}
		  \includegraphics[width = 6cm, keepaspectratio, clip]{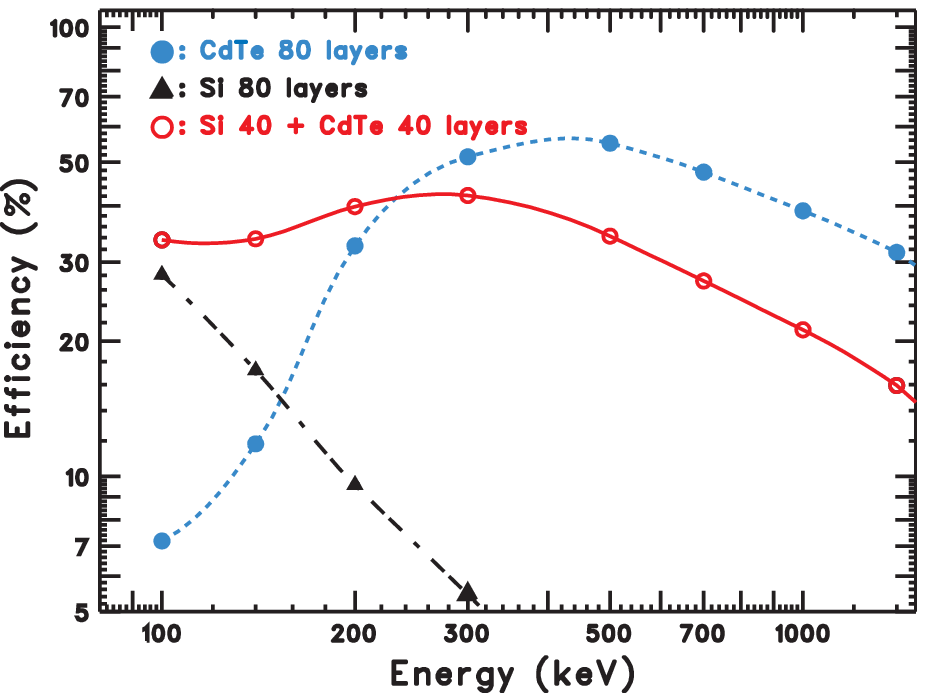}
		\caption[] {Calculated peak detection efficiency for three types of Compton telescopes.  In the simulation, the detector is defined to consist of 80 layers of thin Si or CdTe detector. 
It can be clearly seen that the adoption of CdTe leads to high detection efficiency for a Compton telescope with stacked configuration. [Takahashi, et al, SPIE (2002)]}
		\label{fig: Attn}
	\end{center}
	\end{minipage}
\end{center}
\end{figure}

For energies up to a few MeV, stacking thin CdTe pixel detectors becomes the unique concept to realize a detector with both high energy resolution and high efficiency for gamma-rays \cite{Ref:Takahashi_IEEE2001,Ref:Watanabe_IEEE2002,Ref:Watanabe_IEEE2003}.  As shown in Fig. \ref{fig: Si/CdTe concept}, a stacked CdTe detector is adopted in our Si/CdTe Compton telescope.  
In the stacked detector, several thin CdTe pixel detectors are stacked together and operated as a single detector. By adopting fully-active thin CdTe pixel detector in each layer, we can take advantage of the high energy resolution for gamma-rays from several tens of keV to a few MeV.

Another important merit of utilizing a stacked thin CdTe pixel detector is that it works as an effective Compton telescope above 300 keV, because the stacked detector gives  three-dimensional information of the gamma-ray interaction in the CdTe part.
It should be noted that the cross section for Compton scattering in CdTe becomes higher than that of photo absorption above 300 keV, and the cross section is twice larger than that of Si.  Fig. \ref{fig: Attn} clearly shows how the stacked CdTe works to improve the efficiency for the high energy gamma-rays.  Larger efficiency for the gamma-rays above a hundred keV can be achieved by optimizing the number of layers of CdTe detectors.  Although the angular resolution of a stacked CdTe pixel detector is expected to be worse than that of a Si/CdTe Compton telescope due to the larger Doppler broadening effect \cite{Ref:Zoglauer}, the high efficiency and high energy resolution are attractive for the detection of gamma-rays up to a few MeV.

\section{Experimental Setup} 
In order to operate a stacked CdTe pixel detector as a Compton telescope, 
not only high resolution CdTe pixel detector but also an individual readout system is essential. 
In this section, details of the CdTe pixel detector and the detector system are described.

\subsection{High Resolution 8 $\times$ 8 CdTe Pixel Detectors}

\begin{figure}[]
  	\begin{center}
		  \includegraphics[width = 10cm, keepaspectratio, clip]{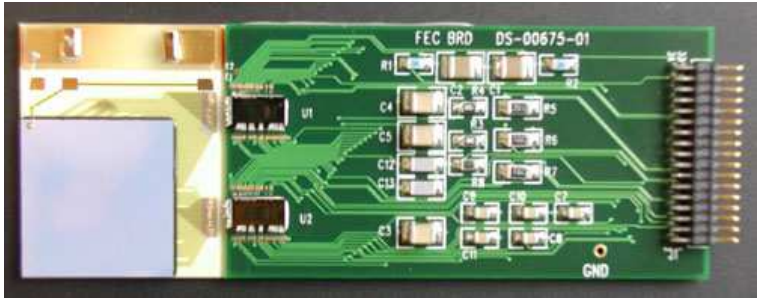}
 \end{center}
		\caption[] {A photo of a 8 $\times$ 8 CdTe pixel detector.  The signal from each pixel is fed into two VA32TAs on the FEC.}
		\label{fig: 8x8 CdTe}
\end{figure}

The 8 $\times$ 8 CdTe pixel detector used in the prototype is based on the Schottky CdTe diode device, utilizing indium as the anode and platinum as the cathode. The CdTe crystal  is manufactured by ACRORAD in Japan with the Traveling Heater Method (THM). It features very low leakage current and high uniformity \cite{Ref:Takahashi_IEEE2002,Ref:Funaki,Ref:KN-NIM2003}. Since position resolution of a few mm is necessary to construct a Compton telescope, we have developed a pixel detector with 64 pixels.  Fig. \ref{fig: 8x8 CdTe} shows the photo of the detector. The detector has dimensions of 18.55 mm $\times$ 18.55 mm and a thickness of 500 \micron. The indium side is used as a common electrode and the platinum side is divided into 8 by 8 pixels.  The pixel size is 2 mm $\times$ 2 mm, and the gap between each pixel is 50 \micron.
A guard ring electrode with a width of 1 mm is attached to reduce leakage current because most of the leakage current is through the detector perimeter. 
Each pixel is connected to a fanout board by bump bonding technology developed in cooperation with
Mitsubishi Heavy Industry in Japan \cite{Ref:Takahashi_IEEE2001}. Since the leakage current of each pixel is as low as 100 pA at a bias
voltage of 500 V even at room temperature, the signal lines are directly connected to the input of readout
electronics, an analog LSI VA32TA \cite{Ref:Tajima2}.  This LSI is developed with IDEAS in Norway, and characterized by its
low noise. The LSIs are read out via a specially designed compact readout system including an ADC and FPGA,
which is controlled with a fast serial interface "Space Wire (IEEE 1355) \cite{Ref:MITANI2}."

\begin{figure}[p]
  	\begin{center}
		  \includegraphics[width = 7cm, keepaspectratio, clip]{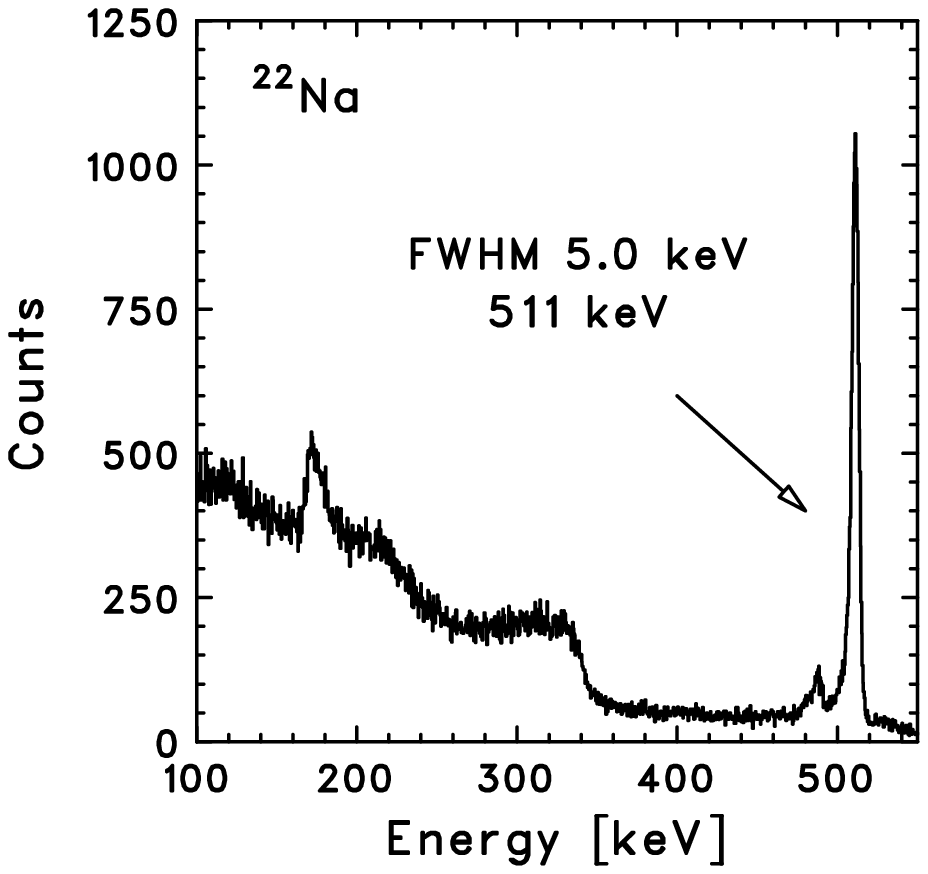}
		  \includegraphics[width = 7cm, keepaspectratio, clip]{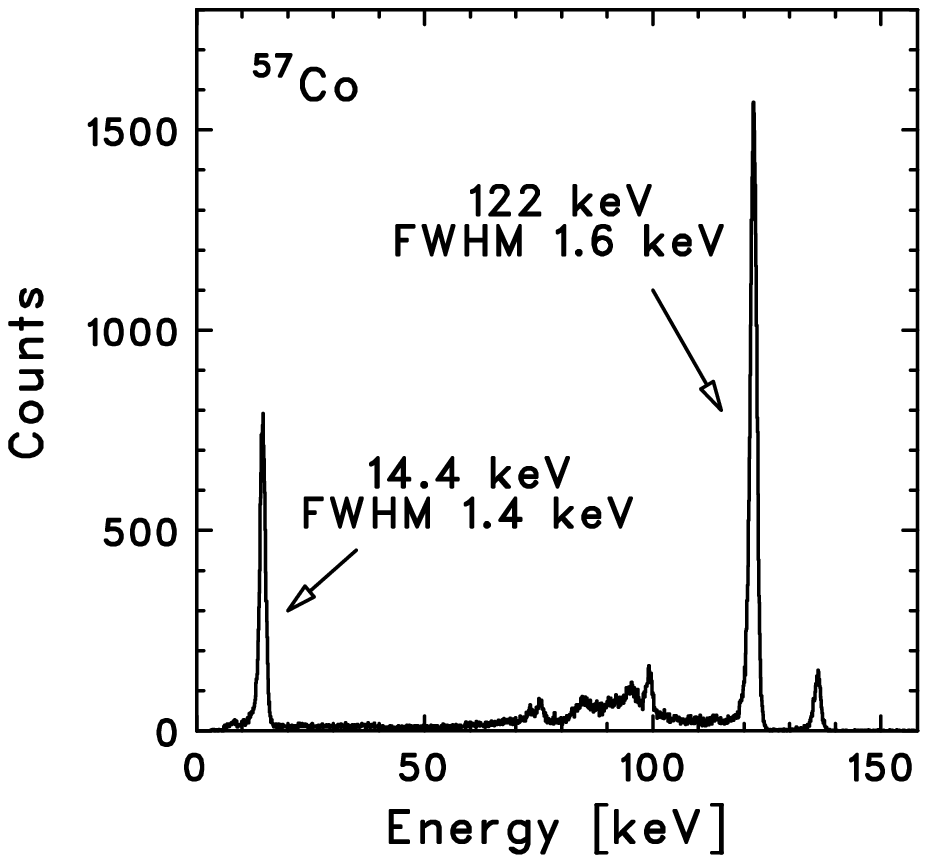}
	 \end{center}
		\caption[] {$^{22}$Na (left) and $^{57}$Co spectra obtained with a 8 $\times$ 8 CdTe pixel detector. The detector is operated at a temperature of $-$20 $^{\circ}$C, with a bias voltage of 1200~V.  Each spectrum is drawn by summing spectra of all 64 channels.  The energy resolution (FWHM) at 511 keV and 122 keV is 5.0~keV and 1.6 keV, respectively. }
		\label{fig: 8x8 CdTe Spectra}
\vspace{1cm}
  	\begin{center}
		  \includegraphics[width = 13.5cm, keepaspectratio, clip]{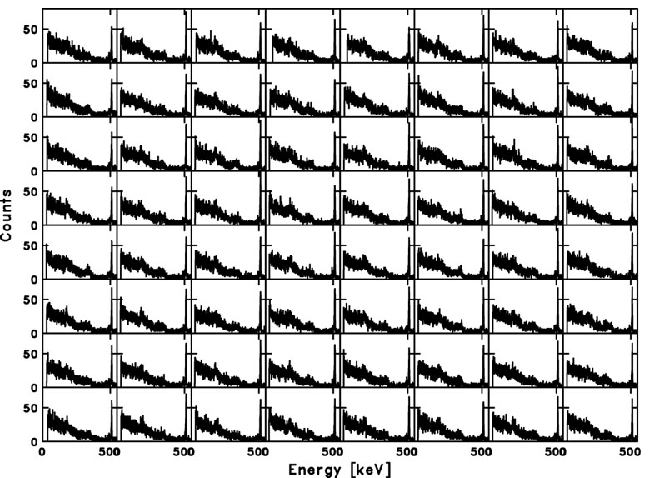}
 \end{center}
		\caption[] { 511 keV spectra of all the 64 pixels.}
		\label{fig: 8x8 CdTe 64ch Spectra}
\end{figure}

High energy resolution of each CdTe pixel detector is essential to achieve both high angular resolution and energy resolution for Compton telescopes.  The low energy threshold of the detector is also important in order to utilize events with small-angle scattering. This is because the deposited energy of a recoiled electron becomes lower as the scattering angle becomes smaller.  
Fig. \ref{fig: 8x8 CdTe Spectra} shows spectra of \Na22{} and \Co57{} with one of the CdTe pixel detectors employed in the prototype stacked CdTe pixel detector.  The spectra are drawn by summing the spectra from all channels after gain correction.  The data is measured at a temperature of $-$20 \degC{} and with a bias voltage of 1200 V.  The energy resolution (FWHM) is 5.0~keV and 1.6~keV for the gamma-rays of 511~keV and 122~keV, respectively.   The energy threshold is measured to be as low as 7~keV.

The homogeneity of the wafer is of particular importance for imaging detectors, because defects in the wafer can deteriorate the energy resolution through positional variation of properties such as charge collection efficiency and leakage current.  As reported in our previous publications \cite{Ref:Takahashi_IEEE2002,Ref:KN-NIM2003}, we have demonstrated the uniform response of large THM-CdTe detectors with an area of 21 mm $\times$ 21 mm.  Fig. \ref{fig: 8x8 CdTe 64ch Spectra} shows spectra of \Na22{} for each channel. The same data as in Fig. \ref{fig: 8x8 CdTe Spectra} (left) is used.  In this detector, all 64 channels are examined to be properly connected, and the detector shows uniform response.

\subsection{Prototype Setup of CdTe Compton Telescope}
  Fig. \ref{fig: Setup} shows the arrangement of the detectors.  Three CdTe pixel detectors are stacked with an interval of 12 mm, while one detector are placed at the side of them.  In order to lower the leakage current and to surpress the so-called ``polarization effect"\cite{Ref:Takahashi_SPIE1998}, the entire detector system was kept at a temperature of $-$20 \degC{}, and the applied bias voltage for the all detectors was set to 700~V.   Under these conditions, we do not see any spectral degradation for a long term operation of a week, as is reported in our paper \cite{Ref:Takahashi_IEEE2002}.
Gamma-ray sources are placed 370 mm above the surface of the first layer of the detector.  A trigger signal is issued from all channels unless it is disabled due to large noise.  When a trigger signal is generated from some channels, all channels of the entire system are read out.

\section{Data Analysis}
\begin{figure}[t]
\begin{center}

	\begin{minipage}[t]{7cm}
	\begin{center}
		  \includegraphics[width = 5cm, keepaspectratio, clip]{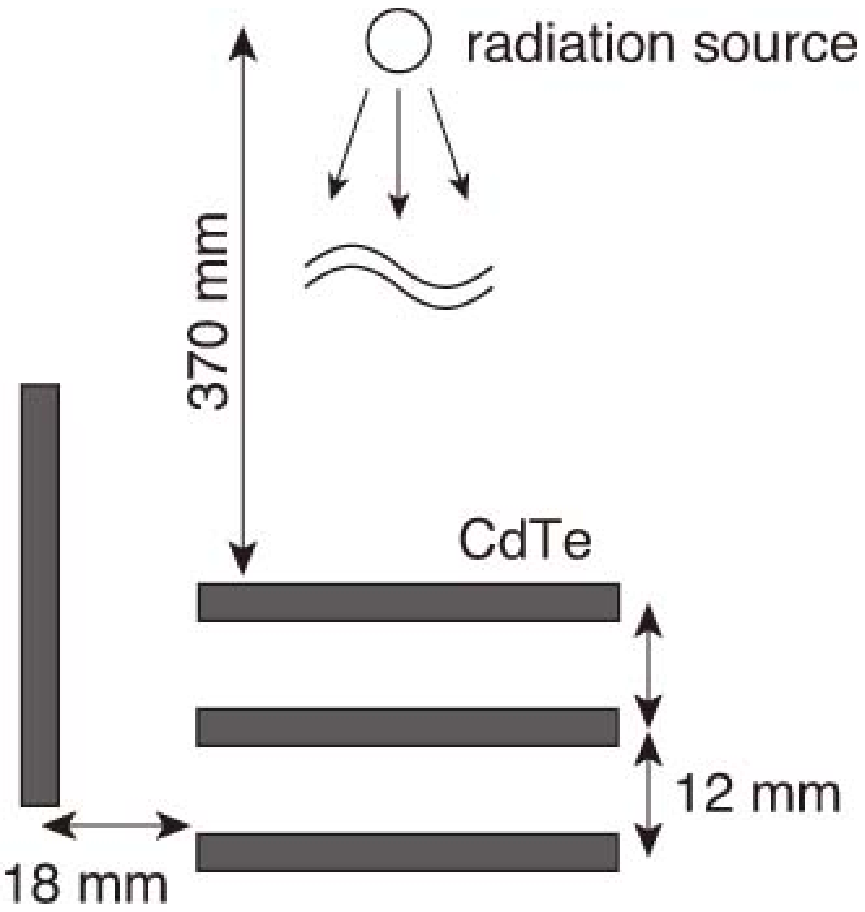}
		\caption[] {Configuration of the prototype stacked CdTe pixel detector.}
		\label{fig: Setup}
	\end{center}
	\end{minipage}
\hspace{0.5cm}
	\begin{minipage}[t]{9cm}
	\begin{center}
		  \includegraphics[width = 7cm, keepaspectratio, clip]{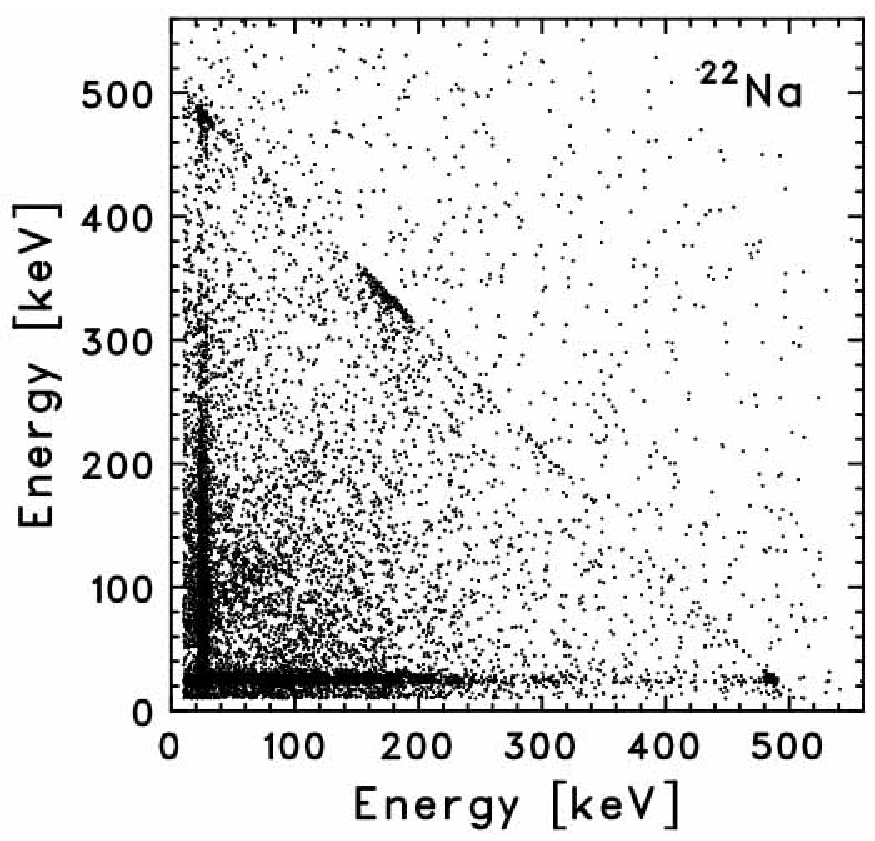}
		\caption[] {Scatter plot of all two-hit events. The plot is obtained by using a \Na22{} gamma-ray source.}
		\label{fig: Scatter Plot}
	\end{center}
	\end{minipage}
\end{center}
\end{figure}

For Compton reconstruction, data reduction is performed as follows. 
First, "two-hit events," one hit in a CdTe pixel detector and one hit in the other detector, were selected from the raw data. Here, one hit means that only one channel has a pulse height above 14 keV.  Fig. \ref{fig: Scatter Plot} shows an example of a scatter plot for the "two-hit events."  This plot is obtained when a \Na22{} radiative isotope is used.  There is a branch which satisfies E $\sim$ 511 keV, which means incident gamma-rays of 511~keV are scattered in one detector and absorbed in the other.  Other noticeable branches are those parallel to the horizontal axis and to the vertical axis between 20~keV and 34~keV, respectively.  They are the events due to fluorescence X-rays from Cd (K$\alpha$: 23.1~keV, K$\beta$: 26.1~keV) and Te (K$\alpha$: 27.4~keV, K$\beta$: 31.0~keV), which escape from the CdTe pixel detectors and are absorbed in the other.  Since they are not the Compton events, we eliminate those events in the analysis.



From each two-hit event, the information of deposited energies and positions are obtained. We name the smaller deposited energy as $E_{1}$, and the other as $E_{2}$ ($E_{1} \leq E_{2}$).
Using these energies, the incident energy ($E_{in}$) and the scattering angle ($\theta_{comp}$) are calculated.  In the calculation, we used:
\begin{eqnarray}
E_{in} & = & E_{1} + E_{2}
\label{eq: Compton Calculation Energy}
\\
\cos \theta_{comp} & = & 1 - \frac{m_{e}c^2} {E_{2}} + \frac{m_{e}c^2}{E_{1}+E_{2}},
\label{eq: Compton Calculation}
\end{eqnarray}
assuming that incident gamma-rays are once scattered and then fully absorbed in the CdTe pixel detector.  Eq. \ref{eq: Compton Calculation} assumes that a incident gamma-ray first deposits its energy ($E_{1}$) by a Compton scattering, and then deposits the rest of the energy ($E_{2}$) by photo absorption. If the energy of incident gamma-rays ($E_{in}$) is less than m$_{e}$c$^{2}$/2 = 255.5 keV, this assumption is valid because the energy of the recoil electron is always less than that of the scattered photon.  If the energy of the incident gamma-rays exceeded 255.5 keV, we calculated two scattering angles by swapping $E_{1}$ and $E_{2}$ in Eq. \ref{eq: Compton Calculation}.  From the calculated scattering angle and the hit positions, Compton cones are drawn on the sky event by event.  The Compton cones are weighted by using the differential cross section of Compton scattering using the Klein-Nishina formula, and the cross section of photo absorption.  We projected the cone at the plane at the distance of 370 mm, and obtained the image of the gamma-ray sources.  

\subsection{Compton Reconstruction of Images}
\begin{figure}[p]
\begin{center}
	
	\begin{minipage}[t]{8cm}
	\begin{center}
	\includegraphics[height=7.4cm, keepaspectratio, clip]{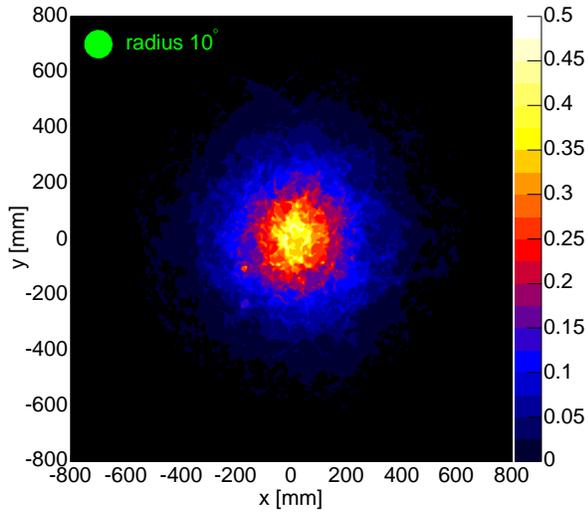}
(a) \Co57{} : 122 keV
	\end{center}
	\end{minipage}
\hspace{0.5cm}
	\begin{minipage}[t]{8cm}
	\begin{center}
		  \includegraphics[height=7.4cm, keepaspectratio, clip]{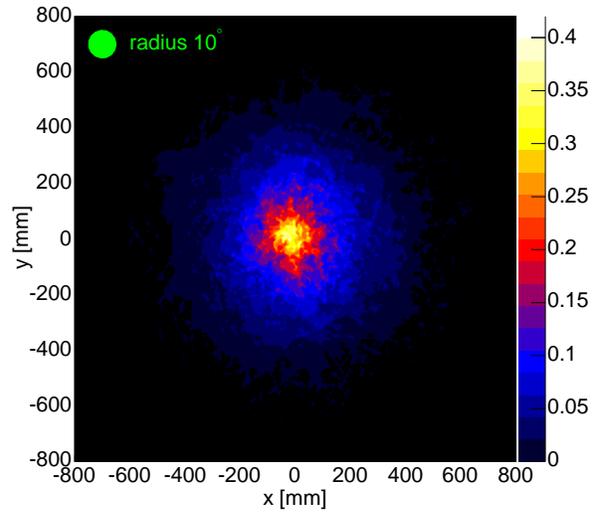}
(b) \Ba133{}: 276, 303, 356, 384 keV
	\end{center}
	\end{minipage}

\vspace{.5cm}

	\begin{minipage}[t]{8cm}
	\begin{center}
	\includegraphics[height=7.4cm, keepaspectratio, clip]{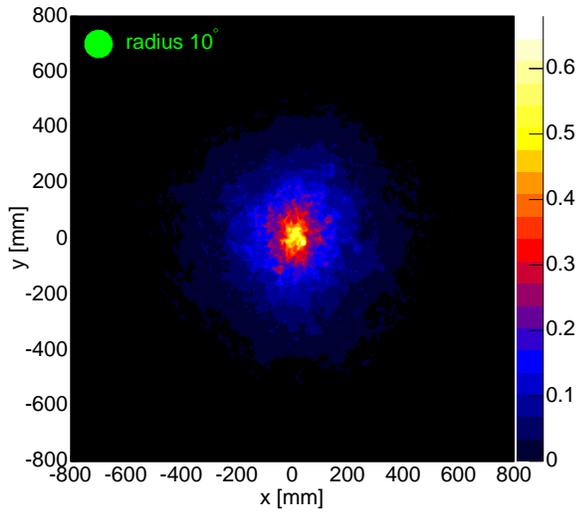}
(c) \Na22{}: 511 keV
	\end{center}
	\end{minipage}
\hspace{0.5cm}
	\begin{minipage}[t]{8cm}
	\begin{center}
		  \includegraphics[height=7.4cm, keepaspectratio, clip]{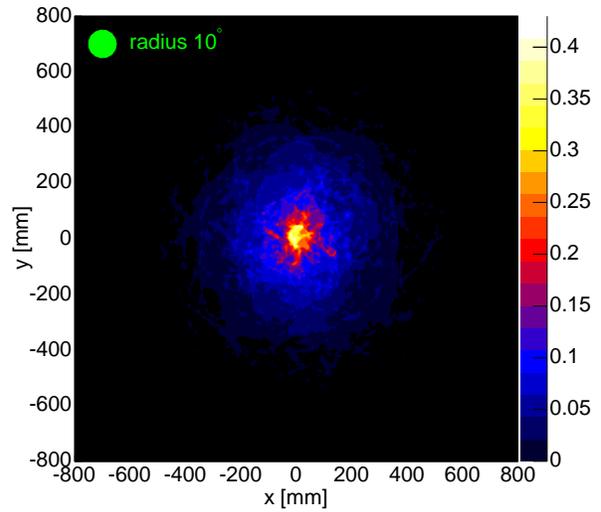}
(d) \Cs137{}: 662 keV
	\end{center}
	\end{minipage}
\caption{Compton reconstructed images. (a) \Co57{} image using 122 keV $\pm$ 5 keV. (b) \Ba133{} image using 270 keV -- 390 keV. (c) \Na22{} image using 500 keV -- 520 keV. (d) \Cs137{} image using 650 keV  -- 675 keV.}
\label{fig: Compton Reconstruction Image}
\end{center}
\end{figure}

Fig. \ref{fig: Compton Reconstruction Image} shows the images obtained with four gamma-ray sources.  A circle with a radius of 10 \degree{} is drawn together with each image as a reference. Fig. \ref{fig: Compton Reconstruction Image} (a) is the image of \Co57{} gamma-rays, by using events within 122 keV $\pm$ 5~keV. The images of \Ba133{}, \Na22{}, and \Cs137{} are also drawn in Fig. \ref{fig: Compton Reconstruction Image} (b) -- (d).  The energy selection is 270 keV -- 390 keV for \Ba133{}, 500 keV -- 520 keV for \Na22{}, and  650 keV  -- 675 keV for \Cs137{}, respectively.  We note that the images are symmetric due to the arrangement of the CdTe pixel detectors being symmetric except for the one at a side.

\subsection{Compton Reconstruction of Spectra}
\begin{figure}[]
\begin{center}
	\begin{minipage}[t]{8.1cm}
	\begin{center}
		 \includegraphics[width = 8cm, keepaspectratio, clip]{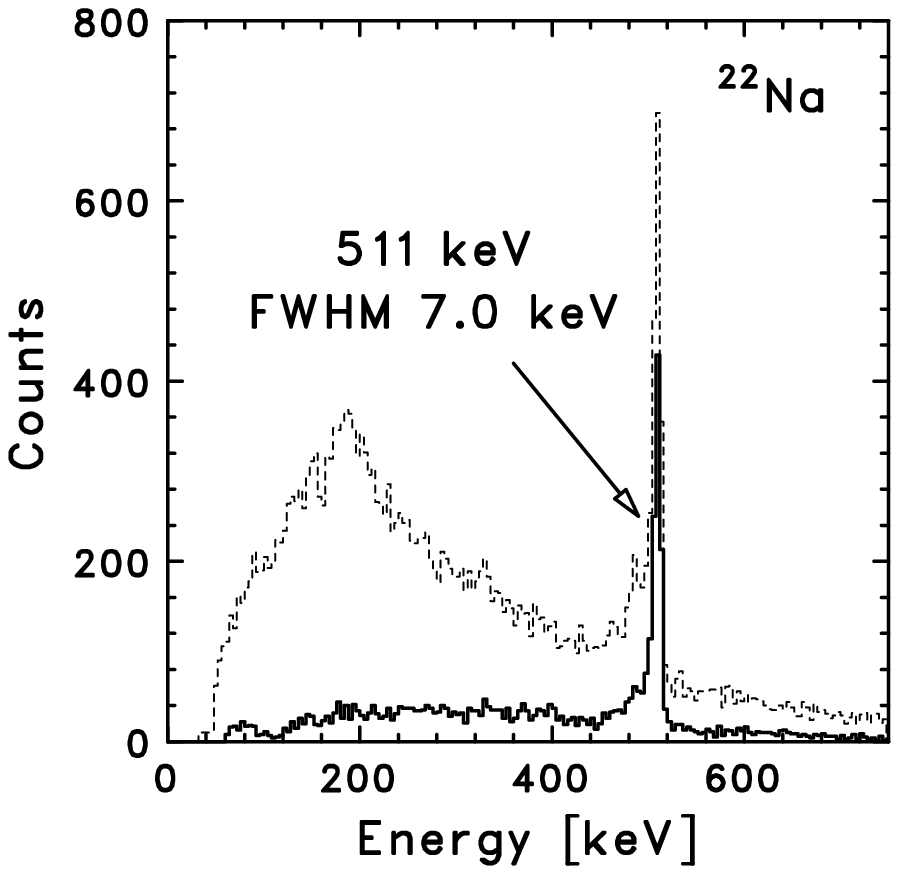}
	\end{center}
	\end{minipage}
\hspace{0.3cm}
	\begin{minipage}[t]{8.1cm}
	\begin{center}
		 \includegraphics[width = 8cm, keepaspectratio, clip]{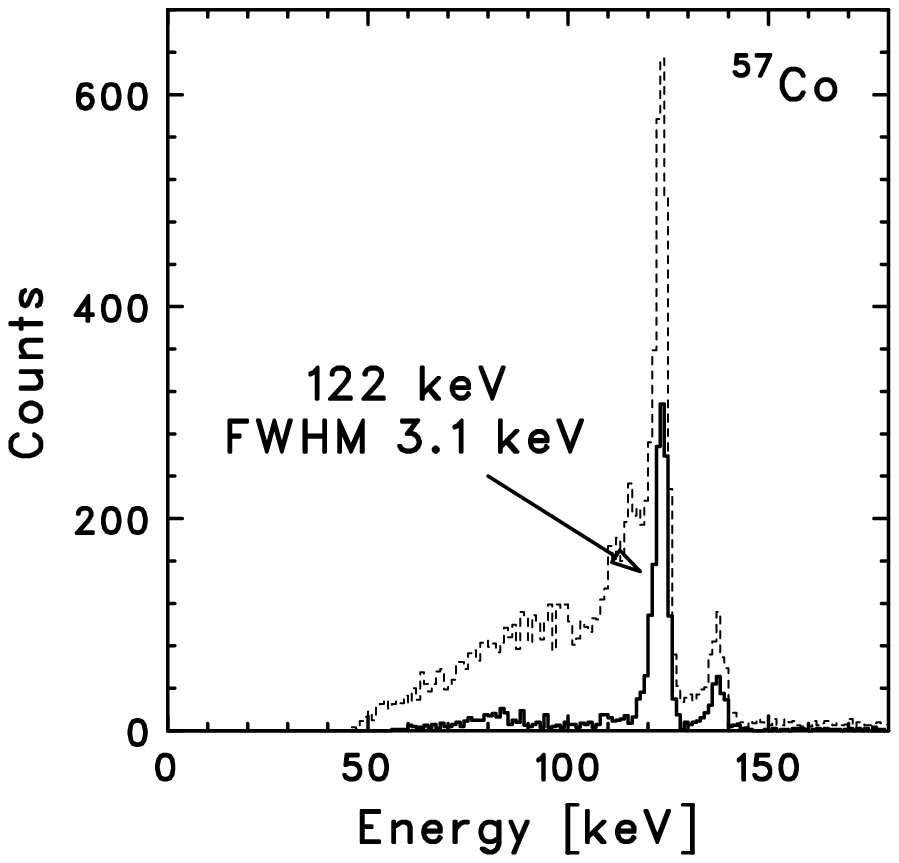}
	\end{center}
	\end{minipage}
\caption{Compton reconstructed spectra of \Na22{} (left) and \Co57{} (right). Dotted lines show the simple sum of two-hit events, while solid lines show the sum of events identified to be emitted from the source by Compton reconstruction.}
\label{fig: Reconstructed Spectra}
\end{center}
\end{figure}

Background rejection using a Compton image is an important feature for a Compton telescope.  
Energy spectra of the incident gamma-rays are obtained by summing deposited energies.  The dotted lines in Fig. \ref{fig: Reconstructed Spectra} shows the spectra which we obtained by simply summing the deposited energies at each detector for all "two-hit events."  In the spectrum obtained with a \Na22{} source, there is a 511 keV gamma-ray line and a large number of scattering components.  The latter components are considered as the events scattered twice in the CdTe pixel detectors, the events   by the gamma-rays scattered in the other materials, and so forth.  Then we performed an event selection using the Compton images.  The solid line spectrum in Fig. \ref{fig: Reconstructed Spectra} is made from the  events which satisfy the condition that the drawn Compton cone corresponds to the direction of the source position.  As shown in the figure, the peak of the 511 keV gamma-rays are enhanced, and the scattering components are reduced.  
The energy resolution (FWHM) is 7.3 keV for 511 keV gamma-rays and 3.1 keV for 122 keV gamma-rays, respectively.

\subsection{Angular Resolution of the CdTe compton Telescope}
\begin{figure}[t]
	\begin{minipage}[t]{8.4cm}
	\begin{center}
	  	\begin{center}
		 \includegraphics[width = 7cm, keepaspectratio, clip]{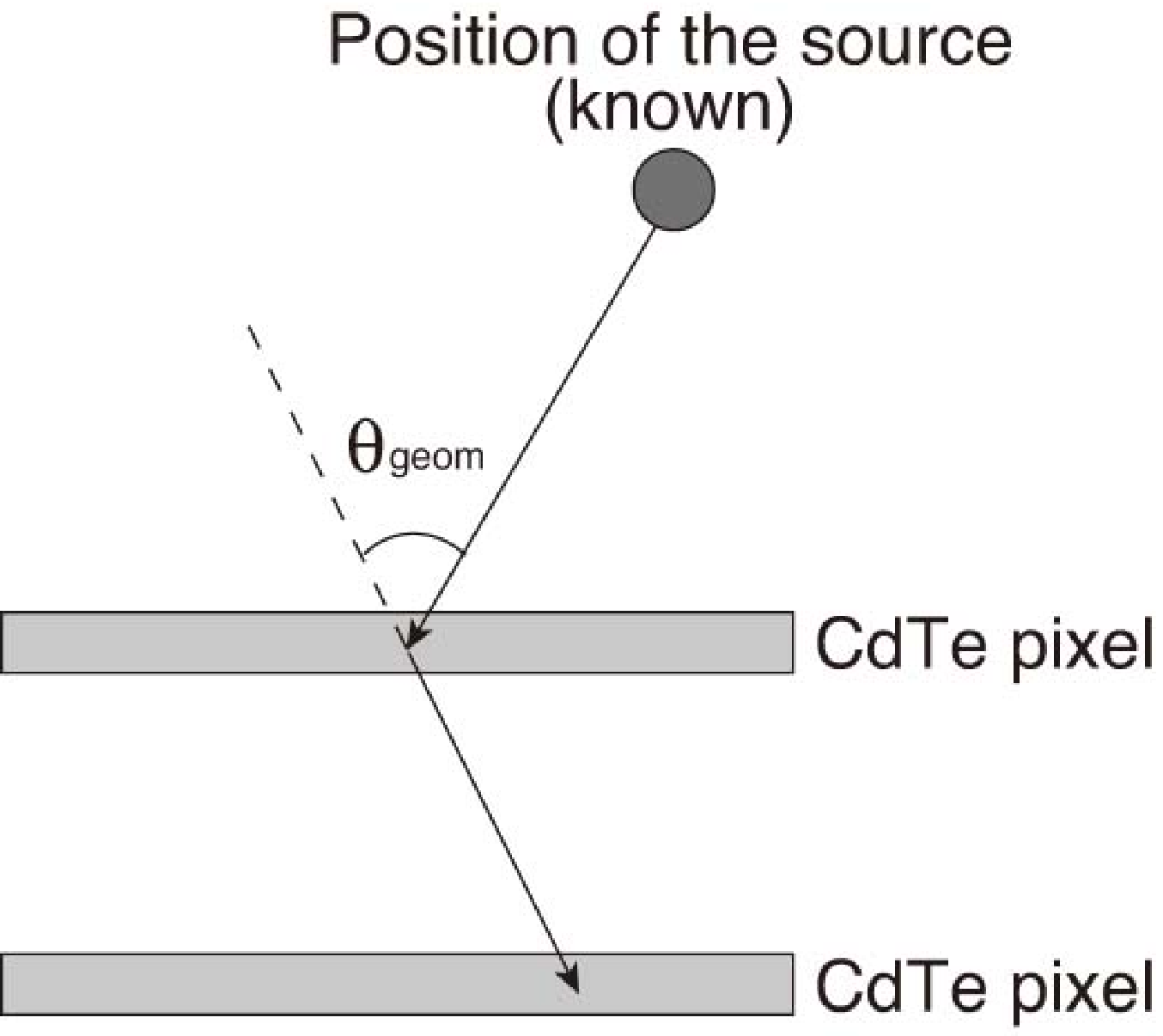}
	 \end{center}
		\caption[] {Definition of $\theta_{geom}$.}
		\label{fig: Theta Geom}
	\end{center}
	\end{minipage}
\hspace{0.5cm}
	\begin{minipage}[t]{8.4cm}
  	\begin{center}
  	\begin{center}
		 \includegraphics[width = 7cm, keepaspectratio, clip]{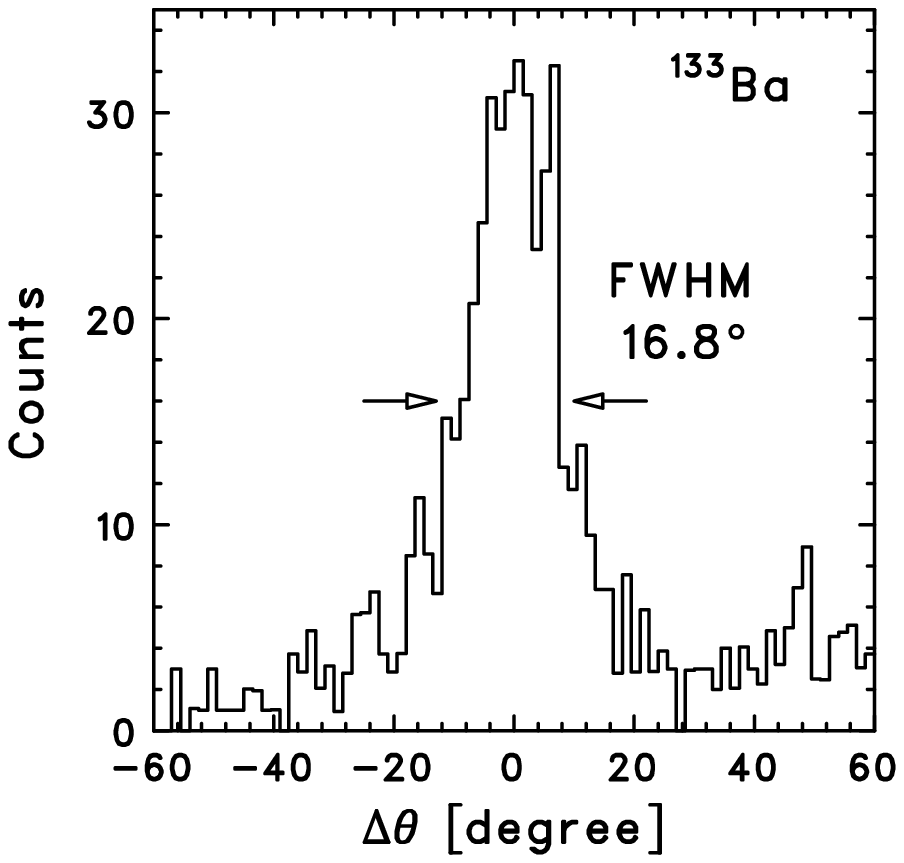}
	 \end{center}
		\caption[] {Distribution of $\theta_{comp} - \theta_{geom}$ for \Ba133 gamma-ray source.}
		\label{fig: Angular Resolution Ba133}
	\end{center}
	\end{minipage}
%
%
  	\begin{center}
	\vspace{0.5cm}
		 \includegraphics[width = 10cm, keepaspectratio, clip]{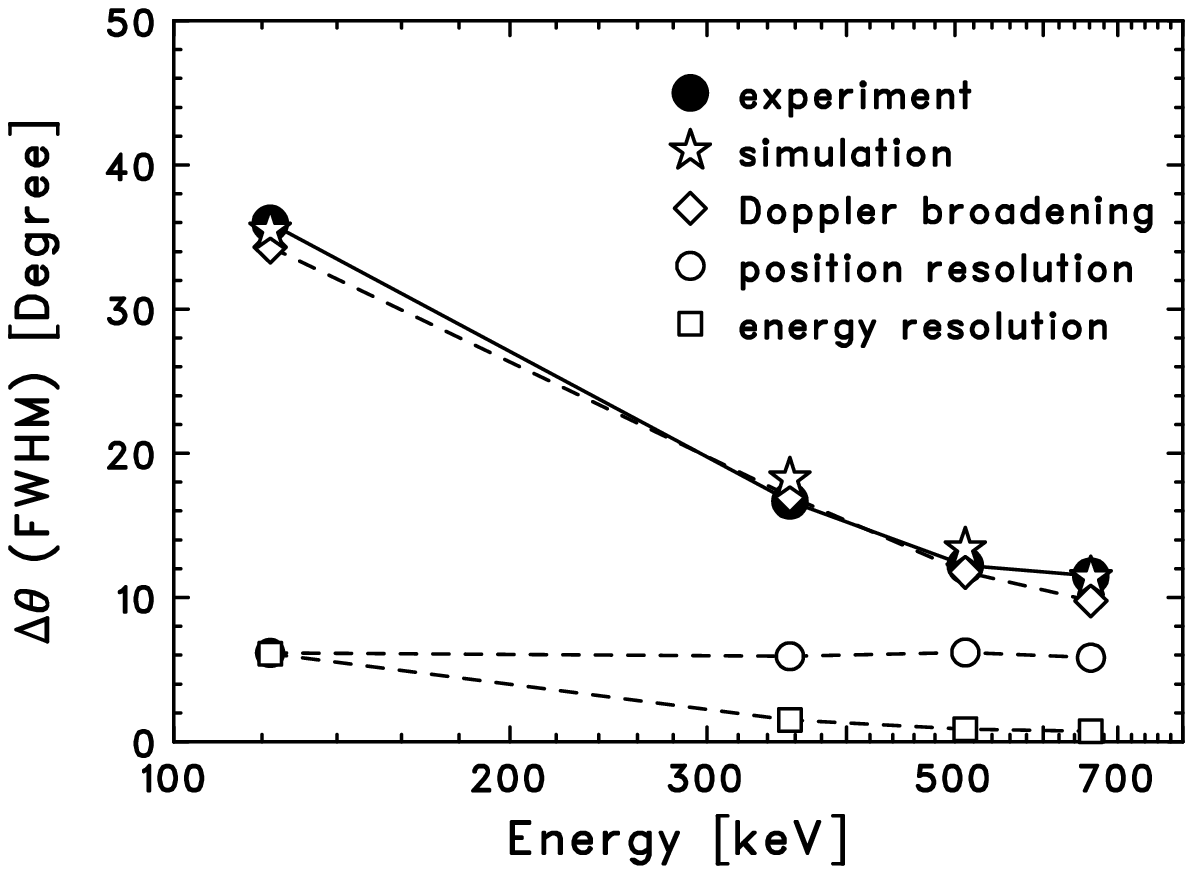}
	 \end{center}
		\caption[] {Relation between incident gamma-ray energy and angular resolution (FWHM).  Filled circles show the experimental data.  Diamonds, open circles, and squares are estimated contributions of Doppler broadening, position resolution, and energy resolution, respectively.  The total values of the three are ploted by stars.}
		\label{fig: Angular Resolution}
\end{figure}

In order to evaluate the angular resolution of our prototype Compton telescope, we compared the calculated scattering angles ($\theta_{comp}$) with those defined by the location of gamma-ray source and hit positions ($\theta_{geom}$) (see Fig. \ref{fig: Theta Geom}). The difference between the two values reflects the angular resolution ($\Delta \theta = \theta_{comp} - \theta_{geom}$) of the Compton telescope.  
Fig. \ref{fig: Angular Resolution Ba133} shows an example of events within 270 keV -- 390 keV for the data obtained with \Ba133{}.  We also investigate the distribution of the angular resolution for the gamma-rays of \Co57{}, \Na22{}, and \Cs137{}.  The FWHM of these distributions ($\Delta \theta_{exp}$), which is the angular resolution of the prototype, are plotted against the incident gamma-ray energies in Fig. \ref{fig: Angular Resolution}.  The angular resolution becomes better as the incident gamma-ray energy become higher.  The obtained angular resolution is 35.9 \degree{} at 122 keV, and  12.2 \degree{} at 511 keV, respectively.

We estimated the contribution of the position and energy resolution of the CdTe pixel detectors.
The contribution of the position resolution ($\Delta \theta_{pos}$) is estimated by using the actual hit positions obtained 
from the experiment, and the pixel size of the detectors.  To estimate the contribution of the energy resolution ($\Delta \theta_{ene}$), we used the Compton-scattering angles of actual events and smoothed them by  the
energy resolution of the detector.  The estimated values are plotted in Fig. \ref{fig: Angular Resolution}.  
The other contribution to the angular resolution is the effect of the Doppler broadening.  In order to estimate the amount of them ($\Delta \theta_{DB}$), we performed Monte Carlo simulations using the Geant4 package \cite{Ref:Kippen}, and plotted them in the same figure.  The effect of the Doppler broadening becomes smaller as the incident energy becomes higher, which is consistent with the paper \cite{Ref:Zoglauer}.  The total value ($\Delta \theta_{total}$) is also plotted in Fig. \ref{fig: Angular Resolution}, which is calculated from:
\begin{equation}
\Delta \theta_{total} = \sqrt{(\Delta \theta_{pos})^2 + (\Delta \theta_{ene})^2 + (\Delta \theta_{DB})^2 }.
\label{eq: Angular Resolution}
\end{equation}
The experimental values can be explained with the three contributions. The angular resolution is almost limited by the effect of Doppler broadening.

\section{Conclusions}
A Si/CdTe semiconductor Compton telescope is a promising detector for future gamma-ray missions in the energy band from several tens of keV to a few MeV.  In order to achieve higher efficiency, we have verified a concept of a stacked CdTe pixel detector which acts as both a scatterer and
an absorber. The prototype detector consists of three layers of CdTe pixel detectors and one CdTe pixel detector at their side.  With this prototype detector, we succeeded in Compton reconstruction of images and spectra in the energy band from 122 keV to 662 keV.  The energy resolution (FWHM) of reconstructed spectra is 7.3 keV at 511 keV and 3.1 keV at 122 keV, respectively.  The high energy resolution of the reconstructed spectra is due to the high performance of thin CdTe pixel detectors.  The obtained angular resolution is 35.9 \degree{} at 122 keV, and  12.2 \degree{} at 511 keV, which is mainly dominated by the effect of Doppler broadening.

The prototype stacked CdTe pixel detector shows higher efficiency in terms of reconstructed spectra than that of the prototype Si/CdTe Compton telescope, while angular resolution of the stacked CdTe is worse than that of the prototype Si/CdTe Compton telescope below 1 MeV. However, the angular resolution becomes a degree scale, when the gamma-ray energy goes up to a few MeV.  Once the stacked CdTe pixel detector is combined with Si layers, the detector should give us a good performance from
several 10 keV up to several MeV.


 
\end{document}